\documentclass{aa}
\usepackage{graphicx}
\usepackage{natbib}
\usepackage{amssymb}
\bibpunct{(}{)}{;}{a}{}{,}
\begin{document}

\title{Probing the mass loss history of carbon stars using CO line and dust continuum emission\thanks{Based 
on observations with ISO, an ESA
project with instruments funded by ESA Member States (especially the
PI countries: France, Germany, the Netherlands and the United Kingdom)
and with the participation of ISAS and NASA.
Radio data collected with the OSO 20\,m telescope, the SEST, and the JCMT,
have also been used.}}

\author{F. L. Sch{\"o}ier\inst{1} \and N. Ryde\inst{2} \and H. Olofsson\inst{3}}

\institute{Leiden Observatory, PO Box 9513, 2300 RA Leiden, The Netherlands
\and Department of Astronomy, University of Texas, Austin, TX 78712-1083, USA
\and Stockholm Observatory, SCFAB, SE-106 91 Stockholm, Sweden} 

\offprints{F. L. Sch{\"o}ier \\ \email{fredrik@strw.leidenuniv.nl}}

\date{A\&A accepted}

\abstract{
 An extensive modelling of CO line emission from the
 circumstellar envelopes around a number of carbon stars is performed. 
 By combining radio observations and infrared observations obtained by ISO 
 the circumstellar envelope characteristics are probed
 over a large radial range.  In the radiative transfer analysis the
 observational data are consistently reproduced assuming a
 spherically symmetric and smooth wind expanding at a constant
 velocity.  The combined data set gives better determined envelope
 parameters, and puts constraints on the mass loss history of these
 carbon stars.  The importance of dust
 in the excitation of CO is addressed using a radiative transfer
 analysis of the observed continuum emission, and it is found to have only
 minor effects on the derived line intensities.  The analysis of the
 dust emission also puts further constraints on the mass loss rate
 history.  The stars presented here are not likely to have experienced
 any drastic long-term mass loss rate modulations, at least less than a
 factor of $\sim$5, over the past thousands of years. Only three, out of
 nine, carbon stars were observed long enough by ISO to allow a
 detection of CO far-infrared rotational lines.
 \keywords{stars: AGB and post-AGB -- circumstellar matter --
            late-type -- mass-loss -- Infrared: stars}
}

\maketitle

\section{Introduction}
Mass loss is associated with many phases of stellar evolution.  It is
of particular importance during the final evolutionary stage of low to
intermediate mass stars, the asymptotic giant branch (AGB),
when it increases dramatically. Its overall characteristics, 
e.g., magnitude, geometry, and kinematics, are reasonably well 
established, but to
what extent such effects as asymmetry, non-homogeneity, temporal
variability, etc., are important is not known.  
During the AGB-phase, stars in the mass range
about 1.5--4\,M$_{\sun}$ and with 'normal' chemical compositions start
to evolve into carbon stars, i.e., stars with more carbon than oxygen
in their atmospheres, through the dredge-up of freshly synthesized
carbon in He-shell flashes \citep{Straniero97, Busso99}. It appears
that all AGB carbon stars are losing mass at a rate in excess of 
10$^{-8}$\,M$_{\sun}$\,yr$^{-1}$ \citep{Olofsson93a, Schoeier01}.

Approximately 10\% of all carbon stars appear to have been subject to
drastic changes in their mass loss rates over relatively short periods
of time (a few thousand years).  Most notable are the CO radio line
observations of detached circumstellar envelopes \citep{Olofsson96,
Olofsson00, Lindqvist99, Schoeier01}.  The time interval between
these, supposedly reoccuring, mass ejections is estimated to be about
10$^5$\,years, and they are possibly linked with the He-shell flashes
predicted to occur in carbon stars \citep{Olofsson90, Schroeder99, Steffen00}.  
Less dramatic mass loss rate
modulations have recently been observed by \citet{Mauron99, Mauron00}
in the form of a multiple-shell structure around the high mass loss rate carbon
star \object{CW Leo} (or \object{IRC+10216}).  This suggests an isotropic mass
loss mechanism with a variability on a broad range of, relatively short, time scales
$\sim$40$-$800\,yr. 
Moreover, spatially resolved interferometric CO millimetre observations
currently have the potential to trace mass loss rate modulations 
on the order of a factor 2$-$3 down to a time scale of several hundred years
\citep{Bieging01}. 
 For changes of the mass loss rate over longer
periods of time, $\gtrsim$10$^{5}$\,yr, statistical studies are
generally required.  The emerging
picture is that of a mass loss rate which increases gradually with time
during the AGB evolution, and its
maximum attainable value depends on the initial main-sequence mass
\citep{Habing96}.  Observational results like these are crucial
for the understanding of mass loss during late stellar evolution, and
for the possibility to pinpoint the mechanism(s) behind this
phenomenon, which determines the time scale during the final evolutionary
AGB-stage.

\begin{table*}
  \caption{Stellar and circumstellar properties of the sample stars.}
  \label{input}
  \begin{center}
  \begin{tabular}{lllcccccc} \hline
  \noalign{\smallskip}
      & & &
  \multicolumn{1}{c}{$D$}    & 
  \multicolumn{1}{c}{$L_*$}    & 
  \multicolumn{1}{c}{{$\dot{M}$}$^{\mathrm d}$}    & 
  \multicolumn{1}{c}{$v_{\mathrm{e}}$} &
  \multicolumn{1}{c}{$r_{\mathrm{p}}$$^{\mathrm e}$} &
  \multicolumn{1}{c}{$h$$^{\mathrm f}$}  
  \\
   \multicolumn{1}{c}{{Source}$^{\mathrm a}$} &
   \multicolumn{1}{c}{{Alt. name}$^{\mathrm b}$} &
   \multicolumn{1}{c}{{Var. type}$^{\mathrm c}$} & 
   \multicolumn{1}{c}{[pc]} & 
   \multicolumn{1}{c}{[L$_{\sun}$]} & 
   \multicolumn{1}{c}{[M$_{\sun}$yr$^{-1}$]} & 
   \multicolumn{1}{c}{[km\,s$^{-1}$]} &
   \multicolumn{1}{c}{[cm]} &
  \\
  \noalign{\smallskip}
  \hline
  \noalign{\smallskip}
   {\object{V384 Per}}        & & Mira & 560 &8100  &3.5$\times$10$^{-6}$ &          15.0 & 1.4$\times$10$^{17}$ & 0.4\\
   {\object{CW Leo}}          & {\object{IRC+10216}} & Mira & 120 &9600  &1.5$\times$10$^{-5}$ &	    14.5 & 3.7$\times$10$^{17}$ & 1.0\\
   {\object{RW LMi}}	      & {\object{CIT 6}} & SRa  & 440 &9700  &6.0$\times$10$^{-6}$ &	    17.0 & 1.9$\times$10$^{17}$ & 1.4\\
   {\object{Y CVn}}           & & SRb  & 220 &4400  &1.5$\times$10$^{-7}$ &\phantom{0}8.5 & 2.9$\times$10$^{16}$ & 0.2\\
   {\object{IRAS 15194-5115}} & & Mira:& 600 &8800  &1.0$\times$10$^{-5}$ &	    21.5 & 3.2$\times$10$^{17}$ & 1.5\\
   {\object{V Cyg}}           & & Mira & 370 &6200  &1.2$\times$10$^{-6}$ &	    11.5 & 8.5$\times$10$^{16}$ & 1.2\\
   {\object{S Cep}}           & & Mira & 340 &7300  &1.5$\times$10$^{-6}$ &	    22.0 & 7.5$\times$10$^{16}$ & 0.4\\
   {\object{AFGL 3068}}       & & Mira:& 820 &7800  &1.5$\times$10$^{-5}$ &	    14.0 & 3.8$\times$10$^{17}$ & 1.5\\
   {\object{LP And}}	      & {\object{IRC+40540}} & Mira & 630 &9400  &1.5$\times$10$^{-5}$ &	    14.0 & 3.8$\times$10$^{17}$ & 0.7\\
  \noalign{\smallskip}
  \hline
  \end{tabular}
  \end{center}
  \smallskip
  
  \noindent
  $^{\mathrm a}$ Parameters taken from \citet{Schoeier01}
  except for \object{IRAS 15194-5115} and \object{AFGL 3068} which are presented in \citet{Ryde99} and Woods et al.\ (in prep.), respectively.\\
  \noindent
  $^{\mathrm b}$ Other frequently used name.\\
  \noindent
  $^{\mathrm c}$ A colon (:) indicates uncertain classification. \\
  \noindent
  $^{\mathrm d}$ A CO abundance of 1.0$\times$10$^{-3}$ relative to H$_2$ was assumed in deriving the mass loss rates.\\
  \noindent
  $^{\mathrm e}$ Estimated CO photodissociation radius.\\
  \noindent
  $^{\mathrm f}$ Dust-grain heating parameter (see text for details).
\end{table*}

Carbon monoxide, CO, is a good tracer of the molecular gas content, and it
has been extensively used to determine the properties of the circumstellar
envelopes (CSEs) formed by the mass loss, e.g., recently
\citet{Schoeier01} used CO millimeter line observations and a detailed
modelling of the emission to determine mass loss rates for a large
sample of optically bright carbon stars.  The Infrared Space
Observatory (ISO; \cite{Kessler96}) opened a new window for observing
AGB-stars, allowing studies of molecular gas, in their expanding
CSEs, much closer to the central stars than was previously possible using
mainly ground based radio telescopes \citep{Cernicharo96, Ryde99}. 
As an example \citet{Ryde99} studied the high mass loss rate carbon star
\object{IRAS\,15194-5115} in several rotational transitions in the
ground vibrational states of $^{12}$CO and $^{13}$CO. In this way
different radial regions of the CSE were probed, and constraints on the
temporal changes in the wind characteristics, in particular the mass
loss rate, were obtained.

In this paper the method used by \citet{Ryde99} is adopted in a study of a
number of carbon stars observed by ISO. The radiative transfer
analysis is further extended to include also the stellar light reemitted by
the surrounding dust, and to investigate its effect on the excitation
of CO.

   \begin{figure}
   \centering{   
   \includegraphics[width=88mm]{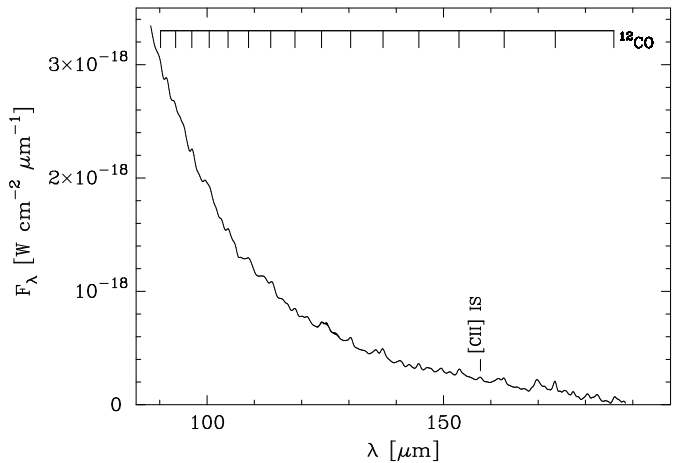}
   \caption{The FIR spectrum of \object{RW LMi}, as observed by ISO,
   showing CO rotational lines in emission superimposed on the continuum
   emitted by the dust present around the star.  All rotational
   transitions within the ground vibrational state in this wavelength
   region (from $J$$=$14$\rightarrow$13 to $J$$=$29$\rightarrow$28) are indicated.  
   In addition, the interstellar [CII] line is marked.}}
   \label{rwlmi_lws}
   \end{figure}

\section{Observations}
Our sample consists of nine carbon stars, of which seven were included
in the large CO radio line survey of carbon stars made by \citet{Olofsson93a}
[see also \citet{Schoeier01} for additional CO observations].
The CO data on \object{IRAS 15194-5115} and \object{AFGL 3068}
are presented in \citet{Ryde99} and 
Nyman et al.\ (in prep.), respectively.  Some of the stellar properties of
the sample sources, as well as the results from detailed radiative
transfer modelling of the CO data [\cite{Schoeier01, Ryde99}; Woods et al.
(in prep.)], are listed in Table~\ref{input}. 

The FIR data were retrieved from the ISO data 
archive\footnote{\tt{http://isowww.estec.esa.nl/}}.  
The observations were carried
out with the Long Wavelength Spectrometer (LWS, Clegg et al.
1996\nocite{Clegg96}), which had a field of view on the sky of
84$\arcsec\times$84$\arcsec$.
%We assume the beam size to be $70\mbox{$\arcsec$}$,
The spectrometer was used in the grating mode (LWS01), covering
the range 90$-$197\,$\mu$m, which provided a mean spectral resolution
element of $\sim$0.7\,$\mu$m.  Since this corresponds to
R$=$$\lambda/\Delta \lambda$$\sim$100$-$250, the circumstellar lines
are far from being resolved.  In what follows, only the
line-integrated flux will be discussed, since no kinematic information
is available.  The spectra were sampled at 0.15\,$\mu$m.

The data reduction was made using the pipeline basic reduction package
OLP (v.\,9.5) and the ISO Spectral Analysis Package (ISAP v.\,2.0). 
The pipeline processing of the data, such as wavelength and flux
calibration, is described in \citet{Swinyard96}; the combined absolute
and systematic uncertainties in the fluxes are on the order of 
$\pm$15--20\%.  The accuracy of the wavelength varies and can be as bad as
$\pm$0.1\,$\mu$m.  The integrated intensities of the emission lines
are measured in the same manner as in \cite{Ryde99}, i.e., a dust
continuum level is estimated and subtracted over a limited wavelength
interval, usually $\sim$10\,$\mu$m.  This procedure introduces
additional, sometimes large, uncertainties in the line fluxes.  The
errors in the fluxes due to the uncertain continuum subtraction are,
on the average, estimated to be of the same order as the calibration uncertainties.  
A total error of  $\pm$25\% in the measured line fluxes is assumed
for all stars.  In the radiative transfer modelling of the CO emission lines observed by
ISO the quality of the fits are good (see Sect.~\ref{results}), 
with reduced $\chi^2$ values $\sim$1, 
indicating that the errors assigned to the fluxes are reasonable.

The final LWS spectra are dominated by thermal radiation due to the
dust present around these objects.  Superimposed on the dust continuum
are molecular emission lines, of which $^{12}$CO, $^{13}$CO, and
H$^{12}$CN lines have been identified in the LWS spectra of \object{CW
Leo} \citep{Cernicharo96} and \object{IRAS 15194-5115} \citep{Ryde99}. 
The LWS spectrum of \object{RW LMi}, where only emission lines arising
from $^{12}$CO have been identified, is presented in
Fig.~\ref{rwlmi_lws}.  The observed fluxes are presented in
Tables~\ref{cwleo}$-$\ref{iras}.  For the rest of the sample stars
presented in Table~\ref{input} only upper limits to the CO FIR line
emission were obtained, Table~\ref{limits}.

\section{Molecular line radiative transfer}
In this section our standard model adopted for the excitation analysis
of the observed CO rotational line emission is presented
\citep{Schoeier01}.  An expansion of the standard model is made to
include a dust component, and  its effect on the CO
excitation is investigated.

\subsection{The standard model}
The observed circumstellar CO line emission is modelled using a
numerical simulation code based on the Monte Carlo method. The code 
solves the non-LTE radiative transfer problem of CO taking into
account 50 rotational levels in each of the fundamental and first
excited vibrational states. 
The CSE is assumed to be spherically symmetric and to expand at a
constant velocity.  The code simultaneously solves the energy balance
and statistical equilibrium equations allowing for a self-consistent
treatment of CO line cooling.  After convergence is reached in the level
populations the radiative transfer equation is solved exactly for the
transitions of interest.  The resulting brightness distributions are
subsequently convolved with the appropriate beams to allow a direct
comparison with the observational data.  See \citet{thesis} and
\citet{Schoeier01} for details on the radiative transfer.
The code has been tested against other radiative transfer codes
for a set of benchmark problems to a high accuracy \citep{Zadelhoff02}.

The recently published collisional rates of CO with H$_2$ by
\citet{Flower01} have been adopted assuming an ortho-to-para ratio of 3.
For temperatures above 400\,K the rates from \citet{Schinke85} 
were used and further extrapolated to include transitions up to $J$$=$50. 
The collisional rates adopted here
differ from those used in the previous modelling of these sources 
\citep{Ryde99,Schoeier01} and accounts for the slightly different envelope
parameters derived in the present analysis.

The size of the CO envelope was fixed to 4$\times$10$^{17}$\,cm and
2$\times$10$^{17}$\,cm for \object{CW Leo} and \object{RW LMi}, respectively.
The adopted sizes reproduce the extents of the CO($J$$=$1$\rightarrow$0) emission
as mapped by the IRAM PdB interferometer for these two sources \citep{Neri98}.
The radial brightness
distribution of the CO($J$$=$1$\rightarrow$0) emission towards 
\object{RW LMi} shows signs of a second, weak and more extended, component 
possibly produced by a modulation of the mass loss rate \citep{Schoeier01}. 
Inclusion of this component in the modelling only marginally affects the derived
CO($J$$=$1$\rightarrow$0) line intensities. Transitions involving higher $J$-levels
are unaffected and this second component is not further considered here.
For \object{IRAS 15194-5115} no interferometric observations exist at present
and the, marginally resolved, CO($J$$=$1$\rightarrow$0) single-dish map
presented by \citet{Nyman93} was used. 
A CO envelope size of $\sim$2$\times$10$^{17}$\,cm, reproduces the 
observed radial brightness distribution well. In the parameter space investigated 
here only the CO($J$$=$1$\rightarrow$0) line intensity is sensitive 
to the adopted envelope size.
However, the effect is generally weak \citep{Schoeier01}.

The standard model includes radiation from a central source
approximated by one or two blackbodies located within the assumed
inner radius of the model CSE. Here we assume that each central source is
described by a single blackbody of the luminosity listed in
Table~\ref{input} and with a temperature of 2200\,K. For the majority of
the sample stars, which have dense CSEs, the line intensities
derived from the model are not sensitive to the adopted description of
the stellar spectrum due to the high line optical depths.  The stellar
photons are typically absorbed within the first few shells in the
model.  Thermal emission from the dust present in the CSE, and the
increase of total optical depth at the line wavelengths, is not properly
treated in the scheme of the standard model.  Dust emission has the
potential of affecting the level populations in the ground vibrational
state, in particular through pumping via excited vibrational states. 
How this is incorporated into the Monte Carlo method is described in
Sect.~\ref{mc_dust}, and its importance for the CO excitation is
discussed in Sect.~\ref{dust}.

In order to put constraints on possible mass loss rate modulations in the
sources where CO line emission is detected by ISO (i.e., \object{CW
Leo}, \object{RW LMi}, and \object{IRAS 15194-5115}), the two
adjustable parameters in the radiative transfer analysis, the mass
loss rate and the $h$-parameter, are varied.  The $h$-parameter is
defined as,
\begin{equation}
h = \left( \frac{\Psi}{0.01} \right)
    \left(\frac{2.0\,\mathrm{g\,cm}^{-3}}{\rho_{\mathrm{gr}}} \right)
    \left(\frac{0.05\,\mu{\mathrm m}}{a_{\mathrm{gr}}} \right), 
\end{equation} 
where $\Psi$ is the dust-to-gas mass loss rate ratio, $\rho_{\mathrm{gr}}$ the mass
density of a dust grain, and $a_{\mathrm{gr}}$ the size of a dust grain. 
The normalized values are the ones used by \citet{Schoeier01} 
to fit the CO millimetre and sub-millimetre line emission of
\object{CW Leo} using the standard model, i.e., $h$$=$1 for this
object.  The $h$-parameter enters into the expression for the heating
term due to the momentum transfer from dust to gas

\begin{equation}
\label{heating}
H_{\mathrm{dg}} = \frac{3}{8}\rho_{\mathrm H_2}\frac{\Psi}{a_{\mathrm{gr}}\rho_{\mathrm{gr}}}\frac{{v_{\mathrm{dr}}^3}}{1+\frac{v_{\mathrm{dr}}}{v_{\mathrm e}}}, 
\end{equation} 
where $\rho_{\mathrm H_2}$ is the density of molecular hydrogen and $v_{\mathrm{dr}}$ is the drift velocity between the dust and the gas.
The drift velocity is expressed by [see \citet{Schoeier01} and
references therein]
\begin{equation}
\label{drift}
v_{\mathrm{dr}} = v_{\mathrm d} - v_{\mathrm e} = \left ( \frac{L v_{\mathrm e} Q}{\dot{M} c} \right )
^{1/2},
\end{equation} 
where $Q$ is the flux averaged momentum transfer efficiency of the dust
assumed to have a value of 0.03 \citep{Schoeier01}. 
This is the
dominant heating process in the CSE, and it is therefore important in
determining the kinetic temperature of the gas.

The best fit model is found by minimizing the chi-squared statistic
\begin{equation}
\label{chi2_sum}
\chi^2 = \sum^N_{i=1} \left [ \frac{(F_{\mathrm{mod},i}-F_{\mathrm{obs},i})}{\sigma_i}\right ]^2, 
\end{equation} 
where $F$ is the flux and $\sigma_i$ the uncertainty in the measured
flux of line $i$, and the summation is done over all $N$ independent
observations.  In the $\chi^2$-analysis the ISO fluxes are assumed to
be accurate to within 25\%.  The millimetre observations are better
calibrated with an estimated uncertainty within 15\% for these strong
CO sources, which are often used as standard sources at radio
telescopes.

\subsection{Addition of a dust component}
\label{mc_dust}
The addition of a dust component in the Monte Carlo scheme is
straightforward.  Scattering is of no importance at the wavelengths of
interest here, and only emission and absorption by the dust particles
are considered.  The dust is assumed to be closely coupled to the
radiation locally, and the particles emit thermal radiation, described
by the dust temperature $T_{\mathrm d}(r)$, according to Kirchhoff's
law.  
The number of model photons emitted per second by the dust within a small volume 
of the envelope, ${\Delta}V$, and over a small frequency interval, $\Delta\nu$, 
can then be written as 
\begin{equation}
N_{\mathrm d}(r) = \frac{4\pi}{h\nu} \kappa_{\nu} \rho_{\mathrm d}(r) B_{\nu}[T_{\mathrm d}(r)] 
                   {\Delta}V(r) \Delta\nu,
\end{equation} 
where $\kappa_{\nu}$ is the dust opacity per unit mass, $\rho_{\mathrm d}$
the dust mass density, and $B_{\nu}$ the Planck function.
Typically the frequency passband is $\sim$3$\times$$v_{\mathrm e}$ and centered on the line rest frequency
$\nu_0$.
The dust temperature structure is obtained from the detailed dust
radiative transfer modelling described in Sect.~\ref{dust_mod}.  The
model photons emitted by the dust are released together with the other model
photons.  The additional opacity provided by the dust, Eq.~\ref{tau},
is added to the line optical depth.

Moreover, since, in general, the dust and the gas have different
temperatures heat will be exchanged during a collision.  The treatment of this process,
which has implications only for the gas, is adopted from
\citet{Groenewegen94} and it is included as an extra heating/cooling term in
the energy balance equation solved during each iteration by the radiative transfer code.

\section{Dust radiative transfer}
\label{dust_mod}
The FIR spectra of carbon stars generally show bright
thermal radiation due to the large quantity of dust present
in the CSEs around these objects, which absorb and re-emit the stellar
visible and near-infrared radiation.
Dust emission is potentially important for excitation of vibrationally
excited states in molecules.  This may affect the level populations in
the ground vibrational state, whose rotational lines are most often used to
constrain circumstellar models.  In the case of CO the first excited
vibrational state lies at 4.6\,$\mu$m, i.e., close to the peak of the
spectral energy distribution (SED; see Fig.~\ref{iras_sed}).  
A shortcoming of our standard model of radiative transfer in molecular
lines is the lack of a full treatment of the radiative transfer of the
dust.  The gas has no importance for the radiative transfer of the
dust, which allows the dust analysis to be carried out separately in
order to determine, in particular, the dust temperature structure needed
for the more detailed molecular excitation analysis.

   \begin{figure*}
   \centering{ 
   \includegraphics[width=180mm]{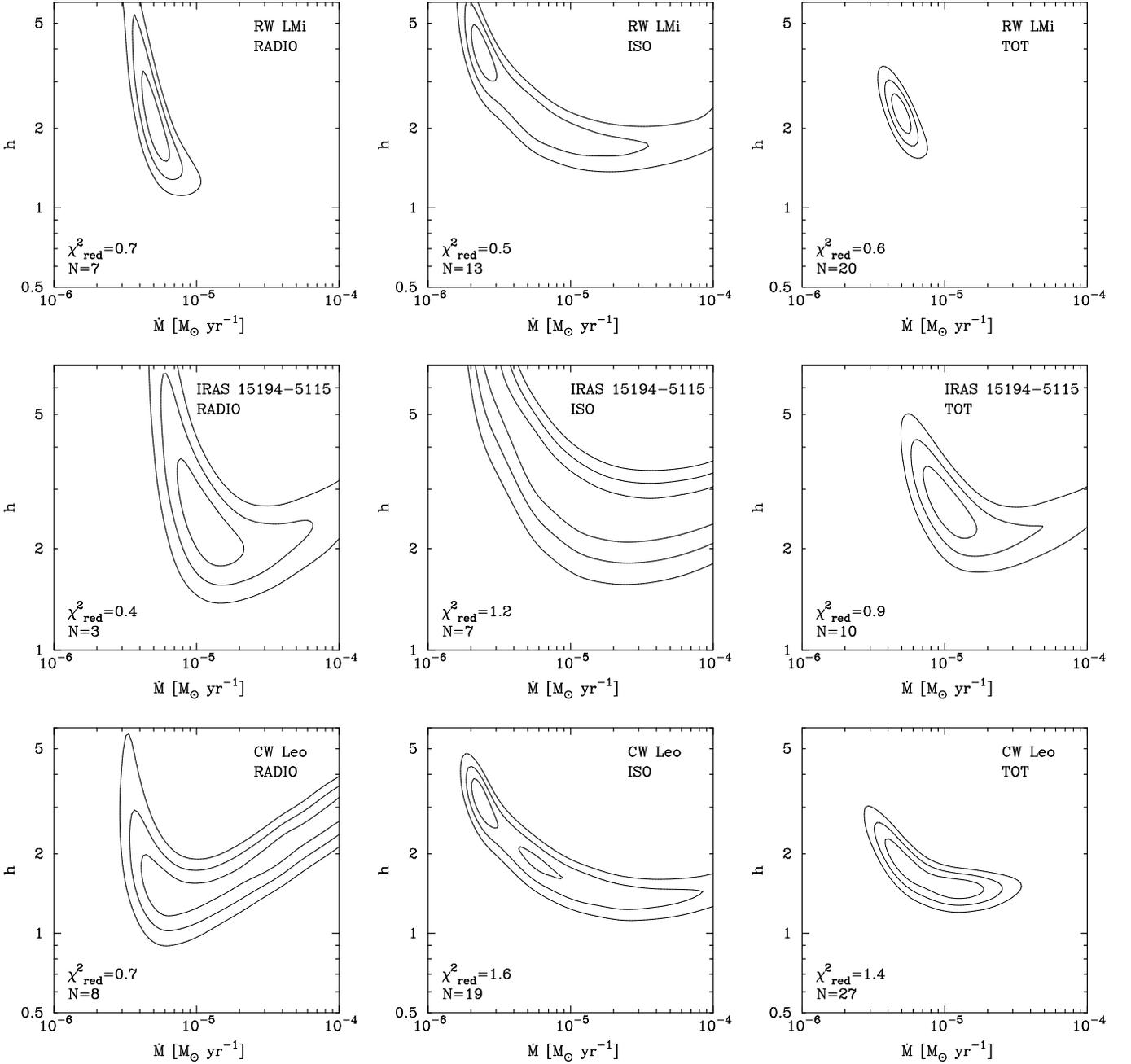}
   \caption{$\chi^2$-maps showing the sensitivity of the CO radiative transfer model,
   using the observed integrated intensities as constraints, to the mass loss rate and
   the $h$-parameter. Contours are drawn at the 1, 2 and 3$\sigma$ levels. The number of observational
   constraints, $N$, are also given. The quality of the best fit models are very good as determined from their
   reduced chi-squared statistics, $\chi^2_{\mathrm{red}}$ $\sim$1.}
   \label{chi2_fig}}
   \end{figure*}

\subsection{Dust modelling}
The radiative transfer problem in a dusty CSE possesses scaling
properties \citep{Ivezic97}.  This fact is put to use in the
publicly available radiative transfer code
DUSTY\footnote{\tt{http://www.pa.uky.edu/\~{ }moshe/dusty/}}
\citep{Ivezic99}, which is adopted here to model the observed continuum
emission.

The most important parameter in the dust radiative transfer modelling
is the dust optical depth
\begin{equation}
\label{tau}
\tau_{\lambda} = \kappa_{\lambda} \int^{r_{\mathrm e}}_{r_{\mathrm i}} \rho_{\mathrm d}(r) dr,
\end{equation} 
where the integration is performed from the inner ($r_{\mathrm i}$) to the
outer ($r_{\mathrm e}$) radius of the CSE.
The dust is assumed to have fully condensated at the inner radius
of the model CSE.

In the dust modelling the same basic assumptions are made as for the gas
modelling, i.e., a spherically symmetric envelope expanding at a
constant velocity.  This results in a dust density structure
where $\rho_{\mathrm d}$$\propto$$r^{-2}$.  
Amorphous carbon dust grains with the optical constants given in
\citet{Suh00} are adopted.  The properties of these grains were shown
to reasonably well reproduce the SEDs in a sample of carbon stars. 
For simplicity, the dust grains are assumed to be of the same
size (a radius, $a_{\mathrm d}$, of 0.1\,$\mu$m), and the same mass
density ($\rho_{\mathrm s}$ equals 2.0\,g\,cm$^{-3}$).  The
corresponding opacities were calculated from the optical constants and
the individual grain properties using standard Mie-theory.
 
In the modelling, where the SED provides the observational constraint,
the dust optical depth specified at 10\,$\mu$m, $\tau_{10}$, and the
dust sublimation temperature, $T_{\mathrm d}(r_{\mathrm i})$, are the
adjustable parameters in the $\chi^2$-analysis.  
The model SED only weakly depends
on the other input parameters which are fixed at reasonable values. 
The effective stellar temperature is set to 2200\,K, i.e., the same value
used in the radiative transfer of the gas.  The size of the CSE is
fixed at $r_{\mathrm e}$/$r_{\mathrm i}$=3000 to ascertain that the whole
CO envelope ($r_\mathrm{ p}$; see Table~\ref{input}) is covered.  The
solution obtained from DUSTY is presented using the relative radius scale
$r/r_{\mathrm i}$.  The $\chi^2$-analysis constrain $\tau_{10}$ and
$T_{\mathrm d}(r_{\mathrm i})$, and a scaling to the adopted luminosity
fixes the absolute radius scale.

\begin{table}
  \caption{Observed and modelled CO FIR rotational lines 
  towards \object{IRC\,+10216}. 
     }
  \label{cwleo}
  \begin{tabular}{rrrrr} \hline
  \noalign{\smallskip}
  \multicolumn{1}{c}{$\lambda_\mathrm{obs}$} &
  \multicolumn{1}{c}{{$F_\mathrm{obs}$}$^{\mathrm a}$} &
  \multicolumn{1}{c}{Transition} &
  \multicolumn{1}{c}{$\lambda_\mathrm{vac}$} &
  \multicolumn{1}{c}{{$F_\mathrm{mod}$}$^{\mathrm b}$} \\
  \multicolumn{1}{c}{[\mbox{$\mu$m}]} &
  \multicolumn{1}{c}{[\mbox{W\,cm$^{-2}$}]} &
   &
  \multicolumn{1}{c}{[\mbox{$\mu$m}]} &
  \multicolumn{1}{c}{[\mbox{W\,cm$^{-2}$}]} \\
  \noalign{\smallskip}
  \hline
  \noalign{\smallskip}
  185.93 &  6.1$\times$10$^{-19}$\phantom{:} &  CO(14-13) & 185.999 & 4.7$\times$10$^{-19}$ \\
  173.60 &  9.0$\times$10$^{-19}$\phantom{:} &  CO(15-14) & 173.631 & 5.0$\times$10$^{-19}$ \\
  162.81 &  4.3$\times$10$^{-19}$\phantom{:} &  CO(16-15) & 162.811 & 5.4$\times$10$^{-19}$ \\
  153.24 &  7.3$\times$10$^{-19}$\phantom{:} &  CO(17-16) & 153.267 & 5.7$\times$10$^{-19}$ \\
  144.70 &  6.6$\times$10$^{-19}$\phantom{:} &  CO(18-17) & 144.784 & 6.1$\times$10$^{-19}$ \\
  137.15 &  7.0$\times$10$^{-19}$\phantom{:} &  CO(19-18) & 137.196 & 6.4$\times$10$^{-19}$ \\
  130.32 &  1.1$\times$10$^{-18}$:                     &	CO(20-19) & 130.369 & 6.8$\times$10$^{-19}$ \\
  124.32 &  8.8$\times$10$^{-19}$\phantom{:} &	CO(21-20) & 124.193 & 7.2$\times$10$^{-19}$ \\
  118.61 &  5.3$\times$10$^{-19}$\phantom{:} &	CO(22-21) & 118.581 & 7.4$\times$10$^{-19}$ \\
  113.34 &  1.3$\times$10$^{-18}$:                     & 	CO(23-22) & 113.458 & 7.6$\times$10$^{-19}$ \\
  108.85 &  5.6$\times$10$^{-19}$\phantom{:} &	CO(24-23) & 108.763 & 7.9$\times$10$^{-19}$ \\
  104.49 &  5.0$\times$10$^{-19}$\phantom{:} &	CO(25-24) & 104.445 & 8.1$\times$10$^{-19}$ \\
  100.35 &  8.9$\times$10$^{-19}$\phantom{:} &	CO(26-25) & 100.461 & 8.3$\times$10$^{-19}$ \\
   96.89 &  1.3$\times$10$^{-18}$:           &	CO(27-26) &  96.773 & 8.5$\times$10$^{-19}$ \\
   93.52 &  7.7$\times$10$^{-19}$\phantom{:} &	CO(28-27) &  93.349 & 8.6$\times$10$^{-19}$ \\
   90.14 &  9.1$\times$10$^{-19}$\phantom{:} &	CO(29-28) &  90.163 & 8.8$\times$10$^{-19}$ \\
   87.20 &  1.1$\times$10$^{-18}$:           &	CO(30-29) &  87.190 & 8.9$\times$10$^{-19}$ \\
   84.43 &  9.0$\times$10$^{-19}$\phantom{:} &	CO(31-30) &  84.411 & 8.9$\times$10$^{-19}$ \\
   81.82 &  4.4$\times$10$^{-19}$\phantom{:} &	CO(32-31) &  81.806 & 8.9$\times$10$^{-19}$ \\
   79.44 &  8.7$\times$10$^{-19}$:           &	CO(33-32) &  79.360 & 8.9$\times$10$^{-19}$ \\
   76.99 &  7.6$\times$10$^{-19}$\phantom{:} &	CO(34-33) &  77.059 & 8.9$\times$10$^{-19}$ \\
   74.49 &  9.4$\times$10$^{-19}$\phantom{:} & 	CO(35-34) &  74.890 & 8.9$\times$10$^{-19}$ \\
   72.85 &  1.2$\times$10$^{-18}$\phantom{:} &	CO(36-35) &  72.843 & 8.8$\times$10$^{-19}$ \\
   70.98 &  1.1$\times$10$^{-18}$\phantom{:} &	CO(37-36) &  70.907 & 8.6$\times$10$^{-19}$ \\
  \noalign{\smallskip}
  \hline
  \end{tabular}
  \smallskip
  
  \noindent
  $^{\mathrm a}$ A colon (:) marks lines with uncertain flux estimates due to
    overlap with an HCN line. These CO lines were not included in the 
    $\chi^2$-analysis.\\
  \noindent
  $^{\mathrm b}$ A mass loss rate of 1.2$\times$10$^{-5}$M$_{\sun}$yr$^{-1}$ and a 
                 $h$-parameter of 1.8 was used in the modelling.
\end{table}
   \begin{figure*} 
   \centering 
   \includegraphics[width=180mm]{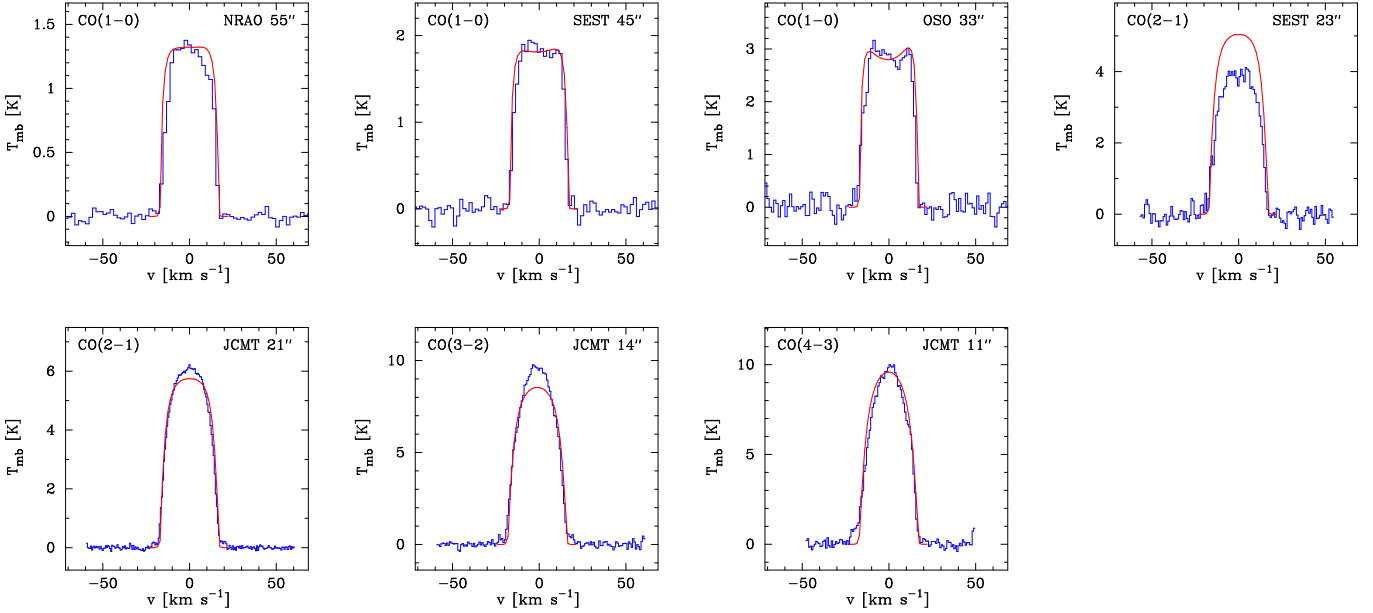}
   \caption{Best fit model (full line) for \object{RW LMi} overlayed on observations (histogram). The CO transition,
   telescope used, and the beamsize, are given for each of the observations.}
   \label{rwlmi_model}
   \end{figure*}

\section{Results and discussion}
\label{results}
In this section the constraints on the CSE characteristics, including the mass loss rate
history, obtained from
the radiative transfer analysis of the CO line emission and the dust
emission are presented.  In addition, the influence of the dust
component on the excitation of CO is investigated.

\subsection{Excitation analysis of CO}
FIR CO line emission was detected by
ISO towards the high mass loss rate carbon stars \object{CW Leo}, \object{RW LMi}, 
and \object{IRAS 15194-5115}, and we start with the analysis of these objects.  
The results of the radiative transfer modelling, varying
$\dot{M}$ and the $h$-parameter, are presented in Fig.~\ref{chi2_fig}
in the form of $\chi^2$-maps with various levels of confidence
indicated.  There are three $\chi^2$-maps per object, showing the
constraints put by the radio and FIR observations individually and
that of the combined full data set.  For low to intermediate mass loss
rate objects it is possible to constrain $\dot{M}$ and $h$
simultaneously from radio observations, provided that multi-transition
observations are available \citep{Schoeier01}.  However, it is evident
that for high mass loss rate objects, i.e., stars with
$\dot{M}$$\gtrsim$10$^{-5}$\,M$_{\sun}$\,yr$^{-1}$, only the
integrated intensities from the observed millimeter lines are
not enough to put good constraints on the mass loss rate.  This is a
consequence of the self-consistent treatment of the CO line cooling,
where an increase of the mass loss rate leads to more cooling which
compensates for the increase in molecular density [see discussion in
\citet{Schoeier01} for details]. 

By combining the radio data with the
additional constraints put by the FIR ISO data it is possible 
to constrain both $\dot{M}$ and $h$ also for the more extreme carbon stars. 
Using the additional information on the excitation conditions contained in the 
shape of the, spectrally resolved, CO radio lines it is occasionally possible to further  
constrain the parameter space. For example, in the case of 
\object{CW Leo} models below $\sim$10$^{-5}$\,M$_{\sun}$\,yr$^{-1}$
have $J$$=$1$\rightarrow$0 line-shapes that are double-peaked indicating 
optically thin resolved emission. This is, however, not observed.
Moreover, models with mass loss rates in excess of $\sim$5$\times$10$^{-5}$\,M$_{\sun}$\,yr$^{-1}$
are too optically thick and produce parabolic line shapes that are too centrally peaked when compared with 
the $J$$=$1$\rightarrow$0 line-shapes.
Estimates of the mass loss rate and the $h$-parameter obtained from the
full set of data are presented in Table~\ref{results}.  
The FIR CO line fluxes
obtained from the best fit models are presented in
Tables~\ref{cwleo}$-$\ref{iras}. In Fig.~\ref{rwlmi_model} the line intensities 
obtained from the best fit model for \object{RW LMi} are overlayed on observations.
Similar plots are presented for \object{IRAS 15194-5115} in \citet{Ryde99} and 
for \object{CW Leo} (radio data only) in \citet{Schoeier01}.

\begin{table}
  \caption{Observed and modelled CO FIR rotational lines 
  towards \object{RW LMi}.}
  \label{rwlmi}
  \begin{tabular}{rrrrr} \hline
  \noalign{\smallskip}
  \multicolumn{1}{c}{$\lambda_\mathrm{obs}$} &
  \multicolumn{1}{c}{{$F_\mathrm{obs}$}} &
  \multicolumn{1}{c}{Transition} &
  \multicolumn{1}{c}{$\lambda_\mathrm{vac}$} &
  \multicolumn{1}{c}{{$F_\mathrm{mod}$}$^{\mathrm a}$} \\
  \multicolumn{1}{c}{[\mbox{$\mu$m}]} &
  \multicolumn{1}{c}{[\mbox{W\,cm$^{-2}$}]} &
   &
  \multicolumn{1}{c}{[\mbox{$\mu$m}]} &
  \multicolumn{1}{c}{[\mbox{W\,cm$^{-2}$}]} \\
  \noalign{\smallskip}
  \hline
  \noalign{\smallskip}
  185.93 &  4.1$\times$10$^{-20}$ &  CO(14-13) & 185.999 & 4.0$\times$10$^{-20}$ \\
  173.62 &  6.5$\times$10$^{-20}$ &  CO(15-14) & 162.811 & 4.5$\times$10$^{-20}$ \\
  162.83 &  4.7$\times$10$^{-20}$ &  CO(16-15) & 162.811 & 4.5$\times$10$^{-20}$ \\
  153.24 &  6.6$\times$10$^{-20}$ &  CO(17-16) & 153.267 & 4.7$\times$10$^{-20}$ \\
  144.70 &  4.4$\times$10$^{-20}$ &  CO(18-17) & 144.784 & 4.9$\times$10$^{-20}$ \\
  130.32 &  6.1$\times$10$^{-20}$ &  CO(20-19) & 130.369 & 5.2$\times$10$^{-20}$ \\
  124.32 &  6.2$\times$10$^{-20}$ &  CO(21-20) & 124.193 & 5.3$\times$10$^{-20}$ \\
  118.61 &  5.5$\times$10$^{-20}$ &  CO(22-21) & 118.581 & 5.5$\times$10$^{-20}$ \\
  108.85 &  5.1$\times$10$^{-20}$ &  CO(24-23) & 108.763 & 5.6$\times$10$^{-20}$ \\
  104.49 &  5.0$\times$10$^{-20}$ &  CO(25-24) & 104.445 & 5.7$\times$10$^{-20}$ \\
   96.89 &  4.6$\times$10$^{-20}$ &  CO(27-26) &  96.773 & 5.8$\times$10$^{-20}$ \\
   93.52 &  5.9$\times$10$^{-20}$ &  CO(28-27) &  93.349 & 5.8$\times$10$^{-20}$ \\
   90.14 &  6.1$\times$10$^{-20}$ &  CO(29-28) &  90.163 & 5.8$\times$10$^{-20}$ \\

  \noalign{\smallskip}
  \hline
  \end{tabular}
  \smallskip
  
  \noindent
  $^{\mathrm a}$ A mass loss rate of 5.0$\times$10$^{-6}$M$_{\sun}$yr$^{-1}$ and a 
  $h$-parameter of 2.3 was used in the modelling.
\end{table}

The quality of the best fit models can be judged from their reduced
$\chi^2$ obtained from
\begin{equation}
\chi^2_{\mathrm{red}} = \frac{\chi^2_{\mathrm{min}}}{N-p},
\end{equation} 
where $p$ is the number of adjustable parameters, in this case two. In all cases the best fit model has 
$\chi^2_{\mathrm{red}}$$\sim$1, indicating a good fit.

\begin{table}
  \caption{Observed and modelled CO FIR rotational lines 
  towards \object{IRAS 15194-5115}.}
  \label{iras}
  \begin{tabular}{rrrrr} \hline
  \noalign{\smallskip}
  \multicolumn{1}{c}{$\lambda_\mathrm{obs}$} &
  \multicolumn{1}{c}{$F_\mathrm{obs}$} &
  \multicolumn{1}{c}{Transition} &
  \multicolumn{1}{c}{$\lambda_\mathrm{vac}$} &
  \multicolumn{1}{c}{{$F_\mathrm{mod}$}$^{\mathrm a}$} \\
  \multicolumn{1}{c}{[\mbox{$\mu$m}]} &
  \multicolumn{1}{c}{[\mbox{W\,cm$^{-2}$}]} &
   &
  \multicolumn{1}{c}{[\mbox{$\mu$m}]} &
  \multicolumn{1}{c}{[\mbox{W\,cm$^{-2}$}]} \\
  \noalign{\smallskip}
  \hline
  \noalign{\smallskip}
  185.99 &  3.0$\times$10$^{-20}$ &  CO(14-13) & 185.999 & 2.6$\times$10$^{-20}$ \\
  173.66 &  3.8$\times$10$^{-20}$ &  CO(15-14) & 162.811 & 2.8$\times$10$^{-20}$ \\
  162.85 &  4.2$\times$10$^{-20}$ &  CO(16-15) & 153.267 & 3.0$\times$10$^{-20}$ \\
  153.38 &  4.0$\times$10$^{-20}$ &  CO(17-16) & 144.784 & 3.1$\times$10$^{-20}$ \\
  144.80 &  3.8$\times$10$^{-20}$ &  CO(18-17) & 137.196 & 3.3$\times$10$^{-20}$ \\
  137.12 &  3.5$\times$10$^{-20}$ &  CO(19-18) & 130.369 & 3.4$\times$10$^{-20}$ \\
  124.10 &  2.8$\times$10$^{-20}$ &  CO(21-20) & 118.581 & 3.7$\times$10$^{-20}$ \\
  \noalign{\smallskip}
  \hline
  \end{tabular}
  \smallskip
  
  \noindent
  $^{\mathrm a}$ A mass loss rate of 1.2$\times$10$^{-5}$M$_{\sun}$yr$^{-1}$
  and a $h$-parameter of 2.8 was used in the modelling.
\end{table}
   \begin{figure*}
   \centering{   
   \includegraphics[width=180mm]{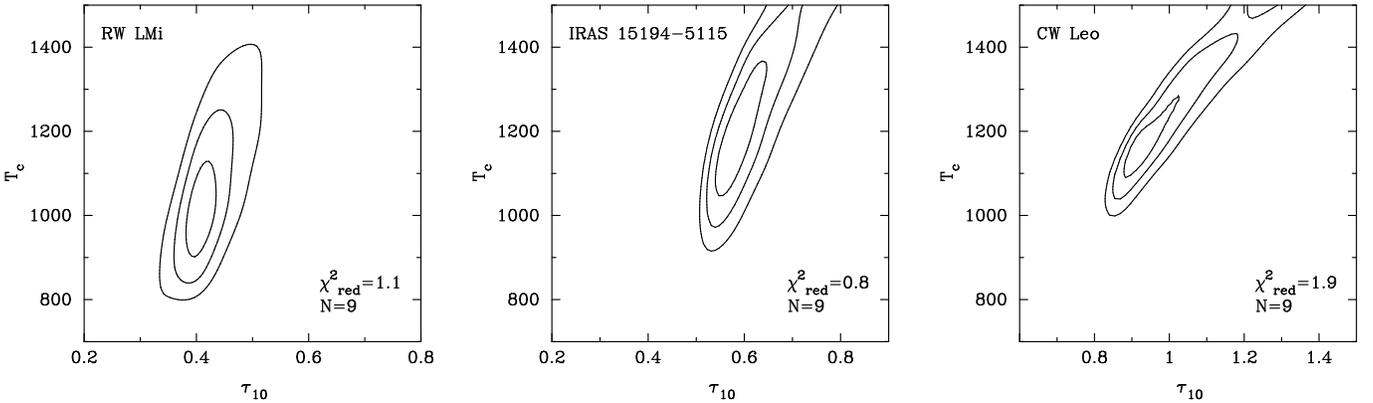}
   \caption{$\chi^2$-maps showing the sensitivity of the dust radiative transfer model,
   using the spectral energy distribution as constraint, to the dust opacity at 10\,$\mu$m ($\tau_{10}$) and
   the dust condensation temperature ($T_{\mathrm c}$). Contours are drawn at the 1, 2 and 3$\sigma$ levels. 
   The number of observational
   constraints, $N$, are also given. The quality of the best fit models are very good as determined from their
   reduced chi-squared statistics, $\chi^2_{\mathrm{red}}$ $\sim$1.} 
   \label{chi2_dust_fig}}
   \end{figure*}
   \begin{figure*}
   \centering{   
   \includegraphics[width=180mm]{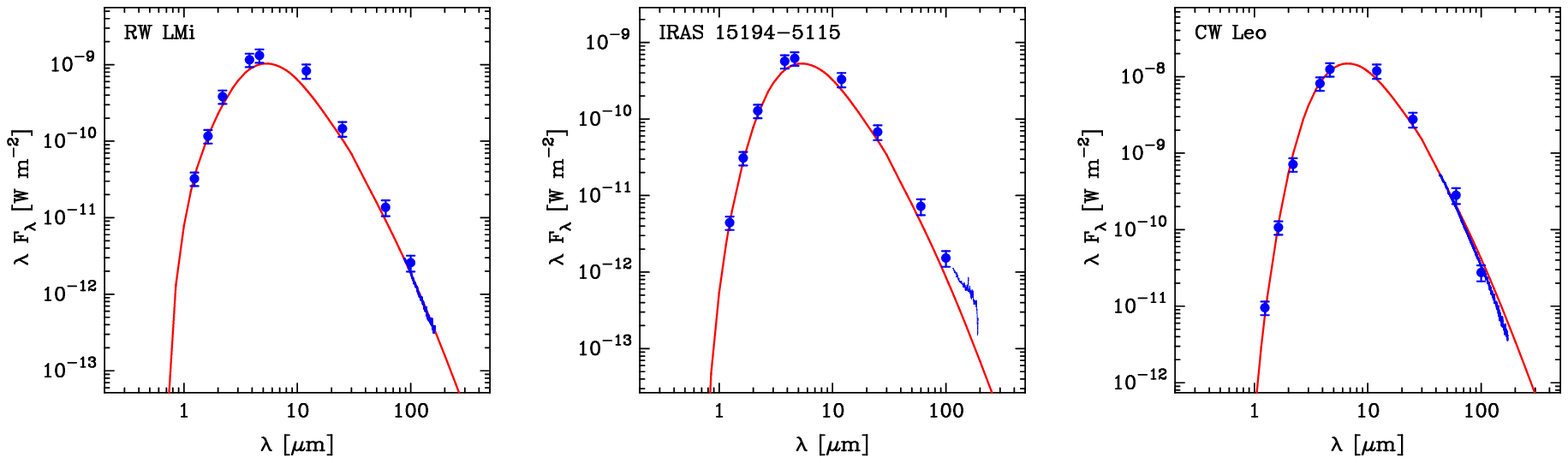}
   \caption{Spectral energy distributions of \object{RW LMi}, \object{IRAS 15194-5115}, and
   \object{CW Leo} obtained from the best fit dust radiative transfer model (full line).  
   Filled circles with error bars
   indicate observed fluxes used in the $\chi^2$-analysis.  Also shown is the ISO FIR spectrum between
   $\sim$100$-$200\,$\mu$m.  The latter was not used to constrain the dust
   radiative transfer modelling.}
   \label{iras_sed}}
   \end{figure*}

A detailed discussion on the excitation of CO, and also the sensitivity of
the results of our standard model to variations of many of the input
parameters, can be found in \citet{Ryde99}, \citet{Schoeier00}, and
\citet{Schoeier01}. 

\subsection{Dust emission}

The dust analysis is only performed for the high mass loss rate carbon stars
\object{CW Leo}, \object{IRAS 15194-5115}, and \object{RW LMi}
where ISO detected CO line emission.
The observational constraints, in the form of SEDs covering the
wavelength range $\sim$1$-$100\,$\mu$m, consist of IRAS fluxes and $JHKLM$-photometry and 
are presented in Fig.~\ref{iras_sed} with
the result from the best fit dust models superimposed. The near-infrared data taken at maximum light 
were obtained from \citet{LeBertre92} for \object{CW Leo} and \object{IRAS 15194-5115}
and from \citet{Taranova99} for \object{RW LMi}.
The fits to the observed SEDs are very good and show no signs of any drastic modulation of the mass loss rate.
In comparison, the presently known detached shell sources show a distinct excess in their 60\,$\mu$m flux
due to a brief, but intense, period of dramatically enhanced mass loss \citep{Olofsson96,
Olofsson00, Lindqvist99, Schoeier01}

The dust optical depth, Eq.~\ref{tau}, can be expressed in terms of the gas mass loss
rate by using the gas-to-dust mass ratio, $\Psi$,
\begin{equation}
\tau_{\lambda} =  \frac{\kappa_{\lambda} \Psi \dot{M}}{4\pi r_{\mathrm i} v_{\mathrm d}},
\end{equation} 
assuming $r_{\mathrm e}$$\gg$$r_{\mathrm i}$.  The optical depth for
the best fit dust model can be translated into a mass loss rate estimate
using the combination of dust parameters as expressed by the $h$-parameter
derived from the gas modelling and the expression for the drift velocity in Eq.~\ref{drift}.
The mass loss rates obtained in this fashion
are presented in Table~\ref{results} and they are all in excellent agreement with those 
presented in \citet{Ivezic95} and \citet{Groenewegen98a}.
The mass loss rates derived from the dust modelling agree well with those estimated from the 
CO line emission, putting further constraints on the mass loss rate history.

In the derivation of the mass loss rates for these extreme carbon stars a constant value of 0.03 was used 
for the flux averaged momentum transfer efficiency $Q$. In a more detailed modelling of the wind dynamics 
$v_{\mathrm e}$, $v_{\mathrm d}$, and $Q$ will have radial
dependences (e.g.,  Habing et~al.\ 1994\nocite{Habing94}; Elitzur \& {Ivezi{\' c}} 2001\nocite{Elitzur01}). 
However, they reach their terminal values close to the star, and
since no direct observational information is available from this
complex region any radial dependence is neglected in the present analysis.
The terminal value of Q will depend on the amount of dust present and its properties. The fact that the 
SEDs for the high mass loss rate objects presented in Fig.~\ref{iras_sed} peak at roughly the same wavelength
($\sim$5\,$\mu$m), indicates that their terminal values of Q are similar. The drift velocity for these sources
is $\sim$3\,km\,s$^{-1}$, i.e., significantly lower than the terminal gas expansion velocity. In this case,
the dominant heating term of the gas, Eq.~\ref{heating}, can be written
\begin{equation}
H_{\mathrm{dg}} \propto \frac{h v^3_{\mathrm{dr}}}{1+\frac{v_{\mathrm{dr}}}{v_{\mathrm e}}} \approx h v^3_{\mathrm{dr}} \propto h Q^{3/2}.
\end{equation}
This means that the uncertainty in the product $hQ^{3/2}$ is the same as derived for $h$ when $Q$ is fixed, i.e., within $\sim$30\%.

\begin{table*}
  \caption{Derived envelope parameters from the radiative transfer analysis.}
  \label{results}
  \begin{center}
  \begin{tabular}{lccccccccc} \hline
  \noalign{\smallskip}
  &
  \multicolumn{3}{c}{{Gas model}$^{\mathrm a}$}    &  &
  \multicolumn{5}{c}{{Dust model}$^{\mathrm b}$}    \\
  \noalign{\smallskip}
  \cline{2-4}
  \cline{6-10}
  \noalign{\smallskip}
      & 
  \multicolumn{1}{c}{$\dot{M}$}    & 
  \multicolumn{1}{c}{$h$}          &
  \multicolumn{1}{c}{$\chi^2_{\mathrm{red}}$}          & &
  \multicolumn{1}{c}{$\tau_{10}$}    &
  \multicolumn{1}{c}{$T_{\mathrm d}(r_{\mathrm i})$}    &
  \multicolumn{1}{c}{$r_{\mathrm i}$}    &
  \multicolumn{1}{c}{{$\dot{M}$}}     &
  \multicolumn{1}{c}{$\chi^2_{\mathrm{red}}$}          
  \\
   \multicolumn{1}{c}{Source} &
   \multicolumn{1}{c}{[M$_{\sun}$yr$^{-1}$]} & & &
   & & 
   \multicolumn{1}{c}{[K]} &
   \multicolumn{1}{c}{[cm]} &
   \multicolumn{1}{c}{[M$_{\sun}$yr$^{-1}$]} 
  \\
  \noalign{\smallskip}
  \hline
  \noalign{\smallskip}
   \object{RW LMi}            &5.0$\pm$1.0$\times$10$^{-6}$ & 2.3$\pm$0.3 & 0.8 & & 0.41$\pm$0.02 & 1050$\pm$100 & 2.2$\times$10$^{14}$ & 3.5$\times$10$^{-6}$ & 1.1\\
   \object{IRAS 15194-5115}   &1.2$\pm$0.5$\times$10$^{-5}$ & 2.8$\pm$0.7 & 0.9 & & 0.59$\pm$0.05 & 1200$\pm$150 & 1.5$\times$10$^{14}$ & 3.5$\times$10$^{-6}$ & 0.8\\
   \object{CW Leo}            &1.2$\pm$0.8$\times$10$^{-5}$ & 1.8$\pm$0.5 & 1.4 & & 0.95$\pm$0.07 & 1200$\pm$100 & 1.9$\times$10$^{14}$ & 8.0$\times$10$^{-6}$ & 1.9\\
  \noalign{\smallskip}
  \hline
  \end{tabular}
  \end{center}
  \smallskip
  
  \noindent
  $^{\mathrm a}$ The uncertainties are based on the 68\% confidence level (1$\sigma$) from Fig.~\ref{chi2_fig} for the combined data set (radio + ISO) and indicates the range 
  of acceptable values for each adjustable parameter. However, note that the 68\% confidence level has a non-rectangular form in the two
  dimensional parameter space.\\
  \noindent
  $^{\mathrm b}$ 
The uncertainties are based on the 68\% confidence level (1$\sigma$) from Fig.~\ref{chi2_dust_fig}  and indicates the range 
  of acceptable values for each adjustable parameter. However, note that the 68\% confidence level has a non-rectangular form in the two
  dimensional parameter space. The uncertainty in the mass loss rate estimate from the dust radiative transfer is $\sim$30-40\% and includes the uncertainty in the adopted $h$-parameter.
\end{table*}
   \begin{figure}
   \centering{   
   \includegraphics[height=75mm, angle=-90]{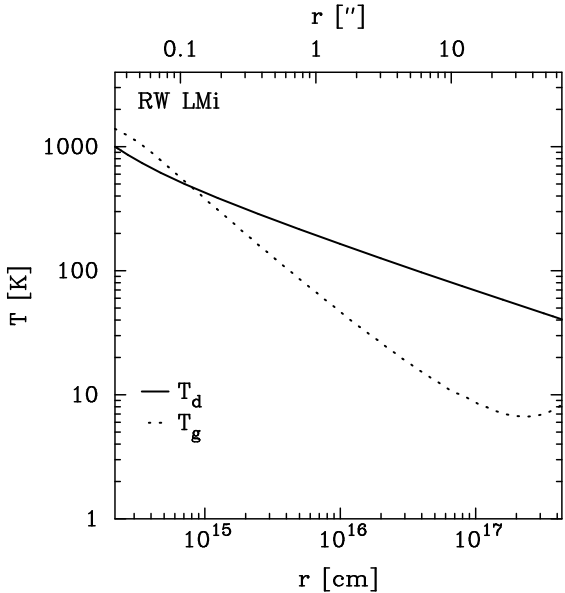}
   \caption{Dust temperature through the circumstellar envelope around \object{RW LMi} (solid line). For 
   $r$$\gtrsim$10$^{15}$\,cm, $T_{\mathrm{d}}(r)$ is well described by a power-law $r^{-0.4}$.
   Also shown for comparison is the kinetic temperature of the gas,
   $T_{\mathrm g}(r)$, obtained from the best fit CO model.}
   \label{iras_temp}}
   \end{figure}

\subsection{Effect of dust emission on the excitation of CO}
\label{dust}
The effect of a dust emission component on the excitation of CO is
tested for \object{RW LMi}.  In Fig.~\ref{iras_temp} the derived dust
and gas temperature structures are plotted.  For
$r$$\gtrsim$10$^{15}$\,cm, the dust temperature is well described by a
power-law falling off as $r^{-0.4}$.  Closer to the star optical
depth effects come into play and $T_{\mathrm{d}}(r)$ decreases faster
with distance.  The dust and gas temperatures are clearly decoupled
throughout the envelope.

\begin{figure}
   \centering{   
   \includegraphics[width=75mm]{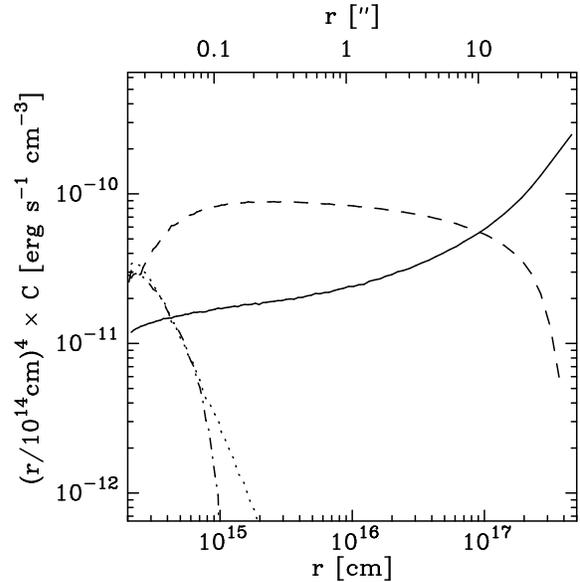}
   \caption{The cooling rates in the CSE around \object{RW LMi} derived from the
   best fit model. 
   The full line represents adiabatic cooling; the dotted line H$_2$
   line cooling; the dashed line the CO line cooling; and the dash-dotted line
   cooling due to heat exchange with the dust.  The cooling rates have
   been multiplied by a factor $\propto$$r^4$ for clarity.  Note that for
   $r$$\gtrsim$10$^{15}$\,cm $T_{\mathrm d}$$>$$T_{\mathrm g}$ and the
   dust will heat the gas.  This extra heating is, however, insignificant
   compared to the heating provided by dust-gas collisions.}
   \label{iras_cool}}
   \end{figure}

In Fig.~\ref{iras_cool} the radial variations of the cooling terms are plotted. 
The cooling rate has been multiplied by a factor $\propto r^4$ for
clarity (the heating due to dust-gas collisions is roughly
proportional to $r^{-4}$).  CO rotational line cooling is the dominant
coolant in the region between $\sim$3$\times$10$^{14}$\,cm and
$\sim$1$\times$10$^{17}$\,cm, where most of the emission from the
observed CO transitions emanate.  Closer to the star cooling due to
H$_2$ line emission and heat exchange between dust and gas becomes
important but very little emission from this part of the CSE
contributes to the line emission observed.  

The excitation analysis shows that only in the innermost parts of the
envelope is the $v$$=$1 state significantly populated, and hence able to
affect the ground-state populations, but this region
is severely diluted in the observational beams, and the derived line
intensities are therefore not notably affected.  In conclusion, the addition of a
dust emission component has only a minor effect on the derived FIR
line intensities, and no effect on the radio line intensities. 
\citet{Kwan77} noted that dust emission have no effect on 
the excitation of the radio lines using the large-velocity-gradient method,
a less detailed treatment than ours of the radiative transfer problem.

\subsection{Mass loss rate modulations}
It is found that the CO radio lines are formed
mainly beyond 5$\times$10$^{16}$\,cm, corresponding to a time scale of
$\ga$10$^3$\,yr, while the FIR lines are formed much closer to the
star, within $\sim$2$\times$10$^{15}$\,cm, which corresponds to
$\la$40\,yr.  Thus, the CO emission modelled here 
traces the mass loss history over
the past several thousand years.  The dust parameters, as represented
by the $h$-parameter, are not likely to have significantly changed during such a
relatively short period of time.  This is further supported by the radiative transfer 
analysis of the dust emission.

From the shape of the $\chi^2$-maps
in Fig.~\ref{chi2_fig} it is clear that the FIR fluxes are degenerate, with respect to the mass loss rate
and the $h$-parameter, to some degree making it hard to put good constraints on any possible mass loss rate modulations. 
An increase in the mass loss rate, and hence the CO density,
can be compensated for by lowering the $h$-parameter and thus the temperature in the envelope producing more
or less the same amount of flux.
In the case of \object{RW LMi} mass loss rate
modulations of a factor of $\sim$2 (1$\sigma$) to $\sim$5 (2$\sigma$) are possible. 
For \object{CW Leo} and \object{IRAS 15194-5115}, 
such constraints can not be obtained from the analysis of the CO emission alone.

The dust emission typically probe the same radial extent of the
envelope as the CO emission.  The quality of the model fits obtained
in the analysis of the SEDs are encouraging (Sect.~\ref{dust}) and
fully consistent with a scenario where the average mass loss rate has
been constant over a large period of time.  The good agreement between
the mass loss rates derived from the dust and the gas modelling
strengthen the conclusion that any long-term mass loss rate
modulations over the past thousands of years are likely to have been
less than a factor of $\sim$5 for these stars.  We conclude, for all
three sources, that the average mass loss rate over the last $\sim$100
years is not substantially different from that over the last few
thousand years.

\begin{table}
  \caption{Observed and modelled FIR fluxes for sample stars where only upper
  limits were obtained.}
  \label{limits}
  \begin{center}
  \begin{tabular}{lcc} \hline
  \noalign{\smallskip}
  Source    & 
  {$F_\mathrm{obs}$}$^{\mathrm a}$ &
  {$F_\mathrm{mod}$}$^{\mathrm b}$ \\
  &
  [W\,cm$^{-2}$] &
  [W\,cm$^{-2}$] \\
  \noalign{\smallskip}
  \hline
  \noalign{\smallskip}
   \object{V384 Per}        & $<$2$\times$10$^{-20}$  & 5.5$\times$10$^{-21}$\\
   \object{Y CVn}           & $<$8$\times$10$^{-21}$  & 2.3$\times$10$^{-21}$\\
   \object{V Cyg}           & $<$4$\times$10$^{-20}$  & 1.7$\times$10$^{-20}$\\
   \object{S Cep}           & $<$3$\times$10$^{-20}$  & 9.3$\times$10$^{-21}$\\
   \object{AFGL 3068}       & $<$2$\times$10$^{-20}$  & 1.4$\times$10$^{-20}$\\
   \object{LP And}          & $<$3$\times$10$^{-20}$  & 1.5$\times$10$^{-20}$\\
  \noalign{\smallskip}
  \hline
  \end{tabular}
  \end{center}
  \smallskip
  
  \noindent
  $^{\mathrm a}$ Upper limit to the CO FIR line emission from observations.\\
  \noindent
  $^{\mathrm b}$ Predicted ISO flux from the model in the CO($J$$=$21$\rightarrow$20) 
  transition using the envelope parameters from Table~\ref{input}.
\end{table}

\subsection{Non detections}
The CSE parameters presented in Table~\ref{input} are used to
calculate the model fluxes in the CO($J$$=$21$\rightarrow$20)
transition for the objects not detected by ISO. The predicted fluxes
are given and compared with the upper limits obtained from the ISO
observations in Table~\ref{limits}.  The model fluxes and the observed
upper limits are all consistent with the scenario of a constant mass
loss rate, or, to be more precise, a mass loss rate which has not
increased significantly over the last 10$^3$ years.  It is also clear
that far too little ISO observing time was spent on each of these six
stars.

\section{Conclusions}

The modelling of CO rotational line emission at millimetre and FIR
wavelengths put constraints on the physical properties of a CSE. Since
the available data probe the full radial extent of the CO envelope
conclusions about temporal changes in the mass loss rate can be drawn. 
For the high mass loss rate objects we find that the FIR data, which
probe the inner regions of the CSE, are consistent with the results
obtained from the radio data.  Under the assumptions of a constant
mass loss rate the combined set of data better constrain the envelope
parameters such as the mass loss rate and the kinetic temperature of the gas.

From the CO data alone, it is generally hard to put good constraints on 
any modulations of the mass loss rate. However, analysis of the 
dust emission put further constraints on the mass loss
rate history, and allows conclusions about its temporal changes to be drawn.
We find that any longer-term mass loss rate
modulations are likely to have been smaller than a factor of $\sim$5
over the past $\la$10$^4$\,yr.  

The role of dust in the excitation of
CO has been investigated and found to be of only minor importance.

\begin{acknowledgements}
We thank Dr.~F.~Kerschbaum for his generous help in providing some of the input data needed
for the analysis.  Support from the ISO Spectrometer Data Centre at MPE Garching, funded by 
DARA under grant 50 QI 9402 3, is acknowledged. 
FLS is supported by the Netherlands Organization for
Scientific Research (NWO) grant 614.041.004. 
NR acknowledge support from the Swedish Foundation for International
Cooperation in Research and Higher Education (STINT).
\end{acknowledgements}

\bibliographystyle{aa}
%\bibliography{refs}

\end{document}